\documentclass[11pt, oneside]{article}   	
\usepackage{geometry}                		
\usepackage{graphicx}				
\usepackage{amssymb}
\usepackage{amsmath}
\usepackage[]{graphicx}
\usepackage{booktabs,siunitx}
\usepackage[flushleft]{threeparttable}
\usepackage[sort,compress]{cite}
\usepackage{float}
\usepackage[caption = false]{subfig}
\usepackage[bf]{caption}
\usepackage{changes}
\begin{document}
\title{Data-Driven Energy Levels Calculation of Neutral Ytterbium (\emph{Z} = 70)\thanks{Project supported by the Fundamental Research Funds for the Central Universities (Grant No. 10822041A2038).}}

\author{Yushu Yu$^{1,2}$, Chen Yang$^{1,2}$\thanks{Corresponding author. E-mail:~yangchen@scu.edu.cn}, and Gang Jiang$^{1,2}$\thanks{Corresponding author. E-mail: gjiang@scu.edu.cn}\\
$^1$Institute of Atomic and Molecular Physics, Sichuan University, \\Chengdu 610065, China\\
  $^{2}$Key Laboratory of High Energy Density Physics and Technology, \\Ministry of Education, Chengdu 610065, China}

\date{\today}
\maketitle

\begin{abstract}
In view of the difficulty in calculating the atomic structure parameters of high-$Z$ elements, the HFR (Hartree-Fock with relativistic corrections) theory in combination with the ridge regression (RR) algorithm rather than the Cowan code's least squares fitting (LSF) method is proposed and applied. By analyzing the energy level structure parameters of the HFR theory and using the fitting experimental energy level extrapolation method, some excited state energy levels of the {Yb~I} ($Z=70$) atom including the $4f$ open shell are calculated. The advantages of the ridge regression algorithm are demonstrated by comparing it with Cowan's least squares results. In addition, the results obtained by the new method are compared with the experimental results and other theoretical results to demonstrate the reliability and accuracy of our approach.
\end{abstract}



\section{Introduction}
The atomic structure parameters of high-$Z$ atoms play fundamental and critical roles in atomic physics. Accurate measurement or calculation of atomic structure parameters is of great significance for understanding the structure of atoms and their ionic states and in some technical fields, such as biological imaging\cite{ning_recent_2022}, medical diagnosis\cite{cisek_wide-field_2017}, atomic light clocks\cite{ludlow_optical_2015}, etc. have a vast range of applications.

In recent years, with the development of photoionization and detection technology, substantial progress has been made in the spectral measurement of complex atomic systems.\cite{eshkabilov_laser_2022,heugel_resonant_2016,sahoo_investigations_2021,xu_study_2018,galindo-uribarri_high_2021,furmann_experimental_2014,kneip_highly_2020,block_direct_2019} However, from the NIST Atomic Spectra Database, there is no systematic experimental data for high-$Z$ elements, especially lanthanides and actinides.\cite{kramida_nist_2022} On the one hand, the high cost of sample preparation and high radioactivity bring much inconvenience in the experimental measurement. On the other hand, although the theoretical calculation method has no cost and safety concerns, with the increase of atomic number, complex electron correlation effects and relativistic effects make the computational complexity of the \emph{ab initio} theoretical method increase exponentially, which is a significant obstacle to using complex atomic systems in fundamental research.

When dealing with a large amount of experimental data, it is appropriate to use machine learning for data mining and modeling research instead of expressing physical equations explicitly in the traditional scientific paradigm. Therefore, data-driven methods are a good way to obtain more atomic structure data, avoid expensive \emph{ab initio} calculations, and improve the accuracy of high-$Z$ atomic data. Nowadays, widely used atomic structure packages can be divided into two categories. The first category is non-relativistic with relativistic corrections packages. For example, the Cowan code\cite{cowan_theoretical_1968} developed by Cowan in 1968 was rather primitive, which adopted the single-configuration radial wave function Hartree-Fock (HF) method. There are also ATSP2K\cite{froese_fischer_mchf_2007} developed by Fischer \emph{et al.} in 2007 using a non-relativistic Multiconfiguration HF (MCHF) method, CIV3\cite{hibbert_civ3_1975} developed by Hibbert, etc. The second category of \emph{ab initio} package is based on the Dirac equation that takes relativistic effects into account, including the GRASP2K\cite{jonsson_grasp2k_2013} \emph{ab initio} package based on the multi-configuration Dirac-Hartree-Fock method published by Grant \emph{et al.} in 2013; the FAC\cite{gu_fac_2008} program with many-body perturbation theory developed by Gu; the MDFGME\cite{MDFGME1,MDFGME2,MDFGME3} program based on the multi-configuration Dirac-Fock matrix element method by Desclaux's group, etc.

 Multi-reference packages, e.g. ATSP2K, GRASP2K, etc. are preferable in calculating the low-level spectra of lighter or specific elements. At the same time, the Cowan code may yield results that are as accurate as or even more accurate in the case of elements containing $nd$ and $nf$ electrons. A major reason is the Cowan code performs the least squares fitting of the \emph{ab initio} calculation results and the existing experimental data on the parameters of atomic structures. As a result, many effects such as the relativistic effect, core polarization, and electron correlation are incorporated into the ab initio calculation. The semi-empirical method has been widely used because of its advantages, such as a relatively simple calculation model and absolute convergence of the iterative process, which significantly improves the calculation accuracy. In 1996, P. Uylings and co-workers\cite{raassen_use_1996} proposed applying the orthogonal operator method in spectral calculations. The orthogonal operator method can be considered an extension and improvement of the traditional least squares fitting (LSF) method, introducing many small parameters to account for specific interactions. The orthogonal operator has many advantages. It uses the same non-relativistic approximation as the Cowan code, with relativistic corrections as a perturbation method. In addition, it takes into account small second-order interactions. However, the orthogonal operator method is only applicable to the $d^n, d^nl, d^{n}l^{n}l^{2}$, and $f^{2}$ configurations so far; for other configurations that partially fill the $f$ shell, the theory remains to be developed. Inspired by the Cowan code, if the current state-of-the-art data-driven machine learning (ML) method is integrated into the calculation of atomic structure parameters, the accuracy of the current atomic structure calculation software can be improved.

Ytterbium ($Z=70$) is one of the typical lanthanide elements, and its neutral atom ground state is $4f^{14}6s^{2}~^1\!S_0$, and two energy level systems are generated about the neutral Yb excited state: The $4f^{14}nln'l'$ configuration excited by $6s$ shell electrons and the $4f^{13}nln'l'n''l''$ configuration level system excited by $4f$ shell electrons. {Yb I} has been recognized by some groups as an excellent optical clock candidate because of its narrow $4f^{14}6s^{2}~^1\!S_0 \rightarrow 4f^{14}6s6p~^3\!P_0$ transition.\cite{hoyt_observation_2005} Energy levels and Landé $g$-factors, as the most basic features of atomic structure spectroscopy, have been studied for a long time. King has measured the spectrum of neutral Yb atoms using an electric furnace in 1931.\cite{king_temperature_1931} Meggers and his colleagues studied atomic spectra of rare earth elements, reporting the wavelengths and estimated intensities of 1791 lines of neutral ytterbium in the region from 2155 to 31308 \si{\angstrom}.\cite{meggers_iii_1942}
 Recently, using microwave and radio frequency resonance methods, F. Niyaz \emph{et al.} observed the transition of ytterbium from $6s(n+3)d~^1\!D_2 \rightarrow 6snl$, $4 \leq l \leq 6$.\cite{niyaz_microwave_2019} In addition to experimental observations, many research groups have developed new theoretical methods to calculate the structural parameters of Yb atoms. V. B. Ternovsky \emph{et al.}\cite{ternovsky_advanced_2018} applied the method of relativistic many-body perturbation and the Dirac-Kohn-Sham zero approximation combined with the generalized theoretical relativistic energy method to calculate the energy and width of the self-ionizing resonance of $4f^{14}7s6p$, $4f^{13}[^2\!F_{7/2}]6s^{2}np[5/2]_2$, $4f^{13}[^7\!F_{7/2}]6s^{2}nf[5/2]_2$. V. A. Dzuba and V. V. Flambaum\cite{dzuba_fast_2019} developed and used an efficient version of the multi-electron configuration interaction method, the fast configuration-interaction method (FCI), to calculate some low states of ytterbium energy levels.

In this work, we further develop a more efficient data-driven method to replace the Cowan code's LSF procedure with the ridge regression (RR) algorithm. Taking the relative energy level of the Yb atom as an example, we compared the energy level results of the partially excited state of the $6s$ shell and the partially excited state of the $4f$ shell calculated by using the LSF and RR models respectively, illustrating the advantages of the RR model in extrapolative prediction, and exploring new methods. In addition, the improved program of RR was used to calculate the energy level and Landé $g$-factor of other parts of the excited state of the {Yb I} atom, and reliable results were obtained.

\section{Calculation methods}
\subsection{Ridge regression algorithm}
Ridge regression\cite{hoerl_ridge_1970} is an improved linear regression algorithm. Unlike the unbiased estimation of linear regression, ridge regression uses $L_2$ regularization to obtain regression coefficients at the expense of losing some information and reducing accuracy, which is more practical and reliable.

For least squares linear regression, its loss function is

\begin{equation}
    J\left( \theta \right) = \sum_{i = 1}^{m}\left( h_{\theta}\left( x^{\left( i \right)} \right) - y^{\left( i \right)} \right)^{2}.
\end{equation}
Using $L_2$ regularization, the loss function should be modified as

\begin{equation}
    J\left( \theta \right) = \sum_{i = 1}^{m}\left( h_{\theta}\left( x^{\left( i \right)} \right) - y^{\left( i \right)} \right)^{2} + \lambda\sum_{j = 0}^{n}\theta_{j}^{2}.
\end{equation}

Using the $L_2$ regularized ridge regression model, taking into account the coefficient $\theta_{j}$ in the prediction function $h_{\theta}\left( x^{\left( i \right)} \right)$, it serves as a penalty mechanism to prevent overfitting of the lines.

\subsection{HFR methods}  
Cowan code's HFR approach optimizes the fine structure parameters to minimize the average deviation between the calculated energy level and the selected experimental value, in order to obtain the required calculated energy level. For an atomic system with nuclear charge number $Z_0$ and electron number $N$, in atomic units, the Hamiltonian of non-relativistic with relativistic correction is

\begin{equation}
    \boldsymbol{H} = - \sum_{i}\nabla_{i}^{2} - \sum_{i}\frac{2Z_{0}}{r_{i}} + \sum_{i > j}\frac{2}{r_{ij}} + \sum_{i}\zeta_{i}\left( r_{i} \right)\boldsymbol{l}_{i}\boldsymbol{s}_{i},
\end{equation}
where the first term is the kinetic energy term of the electron, the second term is the potential energy term of the Coulomb interaction between the nucleus and the electron, the third term is the Coulomb energy term between the electron and the electron, and the fourth term is the spin-orbit interaction term introduced by the relativistic correction.

In theory, solve the Schrödinger equation
\begin{equation}
    H\psi = E\psi,
\end{equation}
the atomic energy levels $E$ and the corresponding atomic wave functions $\psi$ can be obtained. The fitting calculation method is based on the Slater-Condon theory, which expands the unknown atomic wave function $\psi^k$ with a set of known orthonormalized basis functions $\psi_b$ (calculated using Hartree-Fock or several more approximate methods),

\begin{equation}
    \psi^{k} = \sum_{b} Y_{b}^{k}\psi_{b}.
\end{equation}

The atomic wave function is obtained by solving the matrix equation,
\begin{equation}
    \boldsymbol{H}\boldsymbol{Y}^{k} = \boldsymbol{E}^{k}\boldsymbol{Y}^{k}.
\end{equation}

According to the Slater-Condon theory\cite{cowan_theory_1981} used by the HFR method, for the single configuration with the electron occupation number $w_i$ of the orbital $n_{i} l_{i}, \left(n_{1} l_{1}\right)^{w_{1}}\left(n_{2} l_{2}\right)^{w_{2}} \ldots\left(n_{q} l_{q}\right)^{w_{q}}$, its Hamiltonian matrix element is

\begin{align}
    \boldsymbol{H}_{ab} & = \delta_{ab}E_{\text{av}} 
    + \sum_{j = 1}\left\lbrack \sum_{k > 0}\left( f_{k}\left( l_{j}l_{j} \right) \right)_{ab}F^{k}\left( l_{j}l_{j} \right) + \left( d_{j} \right)_{ab}\zeta_{j} \right\rbrack \nonumber\\
    & +  \sum_{i = 1}^{q - 1}\sum_{j = i + 1}^{q}\left[ \sum_{k > 0}\left( f_{k}\left( l_{i}l_{j} \right) \right)_{ab}F^{k}\left( l_{i}l_{j} \right) \right. 
    + \left. \sum_{k}\left( g_{k}\left( l_{i}l_{j} \right) \right)_{ab}G^{k}\left( l_{i}l_{j} \right) \right],\label{Eq:hamiltonian}
\end{align}
where $E_{\text{av}}$ is the configuration average energy, $F^{k}\left(l_{j}l_{j} \right)$ is the Slater direct integral between equivalent electrons, $F^{k}\left( l_{i}l_{j} \right)$ is the Slater direct integral between non-equivalent electrons, $G^{k}\left( l_{i}l_{j} \right)$ is Slater exchange integral between non-identical electrons, $\zeta_{j}$ is the radial integral related to the spin-orbit interaction.

In the HFR method, the one-electron radial wave function for each of the specified electron configurations is first calculated using Hartree-Fock or several more approximate methods, and the $E_{\text{av}}$ for each configuration is obtained along with the Coulomb and spin-orbit interaction \emph{ab initio} value of the integral. Then, an energy matrix is set up for each possible value of $J$, and each matrix is diagonalized to obtain eigenvalues (energy levels) and eigenvectors. Furthermore, relativistic corrections are limited to calculations of mass velocity and Darwin corrections. The effects not considered in other calculations are put into the ridge regression algorithm by treating the integral term in Eq. (\ref{Eq:hamiltonian}) as a variable parameter, and by minimizing its loss function,

\begin{equation}
    R\left( x_{l} \right) = \sum_{k}\left( E^{k} - T^{k} \right)^{2} + \lambda\sum_{l} x_{l}^{2},
\end{equation}
where $E^{k}$ and $T^{k}$ are the calculated values of HFR and experimental energy level, respectively, $x_{l}^{2}$ is the radial integral value, and $\lambda$ is the regularization coefficient. We implemented the ridge regression model to optimize these radial integral parameters, and the average deviation between the calculated energy level and the experimental value is minimized. Therefore, the trained model can achieve a high-precision calculation of the corresponding energy level.

\section{Results and discussion}
In this work, we calculated the relative energy levels and the Landé $g$-factor values of some excited states of {Yb I} atom by using the HFR method improved by the ridge regression model and the least square method used by the Cowan code, respectively, and compared them with the experimental data provided by NIST. In the calculation, the energy level values are relative to the ground state energy level [Xe]$4f^{14}6s^{2}$, and sorted in order from low to high energy levels. Taking the odd-parity configuration $4f^{13}5d6s^{2}$ excited by the $4f$ subshell as an example, we analyzed the advantages of applying the ridge regression algorithm, obtained the connection between the parameters and energy levels, and then applied this law to other highly excited states.

In Table \ref{Tab1}, we list the different methods to calculate the energy level values of all possible spectral items of the odd-parity configuration $4f^{13}5d6s^{2}$, and compare them with the experimental values from NIST respectively, and calculate the mean absolute error (MAE) $R=\sqrt{\sum_i^N{\frac{(E^i-T^i)^2}{N}}}$ for each symmetric block (a symmetric block is a state whose state function differs only by the total angular momentum $J$). The \emph{ab initio} results come from the non-relativistic Hartree-Fock method, with only some of the relativistic effects accounted for as perturbations. The data with asterisks in Table \ref{Tab1} are the data fitted with the experimental values, Fit I--III are the results of fitting with different numbers of experimental values, and the number of experimental values added gradually increases from I--III. It can be clearly seen that adding the experimental energy level effectively improves the accuracy of the energy level calculation, and with the increase of the fitting experimental energy level, the accuracy is higher. In addition, adding experimental energy levels to a certain symmetric block also improves the accuracy of other symmetric blocks. It can be seen from Fit I that when we only add two experimental energy levels to the spectral term (7/2, 1/2), although the Cowan least squares method achieves zero MAE of the symmetric block, however, it affects the results of other symmetric blocks and loses the physical significance. When enough experimental data is added, accurate results can be obtained, and the error with the experimental value is less than 50 \si{\centi\metre^{-1}}. In the case of adding the same less experimental data, comparing the calculation results after adding the ridge regression model program and Cowan code's calculation results, e.g., Fit I and II demonstrate that RR provides more accurate results than Cowan code's LSF procedure, which indicates that the RR algorithm has stronger generalization ability. With less data, it can capture the characteristics of the data more accurately and deliver more accurate results.

\begin{table}[H]
    \caption{{Yb I} $4f^{13}6s^{2}6p$ fitting calculation of energy levels (in \si{\centi\metre^{-1}}). The data with asterisks (*) are the data fitted with the experimental values.}\label{Tab1}
\resizebox{\linewidth}{!}{
\begin{threeparttable}
\begin{tabular}{cccS[table-format=5.1]S[table-format=5.1]S[table-format=5.1]S[table-format=5.1]S[table-format=5.1]S[table-format=5.1]S[table-format=5.1]S[table-format=5.1]}
\toprule
                                         &                                 &                              & \multicolumn{1}{c}{}                                 & \multicolumn{1}{c}{}                                     & \multicolumn{2}{c}{{Fit I}}                                                                   & \multicolumn{2}{c}{{Fit II}}                                                                  & \multicolumn{2}{c}{{Fit III}}                                                              \\ \cline{6-11} 
    {Configuration} & {Term} & {$J$} & \multicolumn{1}{c}{{Expt.}} & \multicolumn{1}{c}{{\emph{Ab initio}}} & \multicolumn{1}{c}{{Cowan code}}              & \multicolumn{1}{c}{{This work}}            & \multicolumn{1}{c}{{Cowan code}}              & \multicolumn{1}{c}{{This work}}            & \multicolumn{1}{c}{{Cowan code}}            & \multicolumn{1}{c}{{This work}}           \\ 
\midrule
$4f^{13}6s^26p$                       & {(7/2, 1/2)}              & {3}                   & 32065.3                                              & 32277.2                                                  & 32065.3*                                          & 32122.1*                                        & 31654.0*                                          & 32065.3*                                         & 32069.0*                                        & 32069.0*                                        \\
                                         &                                 & {4}                   & 32273.6                                              & 32371.3                                                  & 32273.6*                                          & 32216.6*                                         & 33456.0*                                          & 32273.6*                                         & 32280.3*                                        & 32280.3*                                        \\
$R_1$\tnote{a}                             &                                 &                              & \multicolumn{1}{c}{}                                 & \multicolumn{1}{r}{{165.0}}         & \multicolumn{1}{r}{{0.0}}    & \multicolumn{1}{r}{{56.9}}  & \multicolumn{1}{r}{{885.2}}  & \multicolumn{1}{r}{{0.0}}   & \multicolumn{1}{r}{{5.4}}  & \multicolumn{1}{r}{{5.4}}  \\
                                         & {(7/2, 3/2)}              & {5}                   & 35178.8                                              & 34455.3                                                  & 32969.5                                           & 34355.9                                          & 34772.2*                                          & 35165.5*                                        & 35134.8*                                        & 35134.8*                                        \\
                                         &                                 & {2}                   & 35197.0                                              & 34500.5                                                  & 35023.8                                           & 34406.9                                          & 34839.4*                                          & 35210.3*                                        & 35205.4*                                        & 35205.4*                                        \\
                                         &                                 & {3}                   & 35807.5                                              & 34790.8                                                  & 33314.0                                           & 34690.0                                          & 39778.9                                           & 35482.9                                          & 35789.2*                                        & 35789.2*                                        \\
                                         &                                 & {4}                   & 36061.0                                              & 34967.0                                                  & 33471.0                                           & 34867.9                                          & 40314.2                                           & 35747.4                                          & 36104.5*                                        & 36104.5*                                        \\
{$R_2$}                              &                                 &                              & \multicolumn{1}{c}{}                                 & \multicolumn{1}{r}{{899.9}}         & \multicolumn{1}{r}{{2111.7}} & \multicolumn{1}{r}{{996.7}} & \multicolumn{1}{r}{{2922.1}} & \multicolumn{1}{r}{{225.9}} & \multicolumn{1}{r}{{32.5}} & \multicolumn{1}{r}{{32.5}} \\
                                         & {(5/2, 1/2)}              & {3}                   &                                                      & 43213.3                                                  & 46988.0                                           & 43331.2                                          & 42537.0                                           & 42416.3                                          & 42343.4                                         & 42343.4                                         \\
                                         &                                 & {2}                   & 42531.9                                              & 43321.1                                                  & 49249.4                                           & 43443.7                                          & 42196.1*                                          & 42531.9*                                        & 42521.9*                                        & 42521.9*                                        \\
{$R_3$}                              &                                 &                              & \multicolumn{1}{c}{}                                 & \multicolumn{1}{r}{{789.2}}         & \multicolumn{1}{r}{{6717.5}} & \multicolumn{1}{r}{{911.8}} & \multicolumn{1}{r}{{335.8}}  & \multicolumn{1}{r}{{0.0}}   & \multicolumn{1}{r}{{10.0}} & \multicolumn{1}{r}{{10.0}} \\
                                         & {(5/2, 3/2)}              & {3}                   &                                                      & 45833.1                                                  & 48265.1                                           & 46007.1                                          & 49006.2                                           & 45962.9                                          & 46218.5                                         & 46218.5                                         \\
                                         &                                 & {1}                   & 44834.6                                              & 45122.9                                                  & 47625.7                                           & 45297.5                                          & 40650.9                                           & 45258.9                                          & 44877.6*                                        & 44877.6*                                        \\
                                         &                                 & {4}                   & 45497.6                                              & 45434.4                                                  & 47998.0                                           & 45610.6                                          & 43712.0                                           & 45728.3                                          & 45473.3*                                        & 45473.3*                                        \\
                                         &                                 & {2}                   & 45931.9                                              & 45654.2                                                  & 48044.8                                           & 45828.1                                          & 48632.9                                           & 45776.1                                          & 45905.1*                                        & 45905.1*                                        \\
{$R_4$}                              &                                 &                              & \multicolumn{1}{c}{}                                 & \multicolumn{1}{r}{{234.0}}         & \multicolumn{1}{r}{{2483.7}} & \multicolumn{1}{r}{{281.6}} & \multicolumn{1}{r}{{3054.4}} & \multicolumn{1}{r}{{293.0}} & \multicolumn{1}{r}{{32.4}} & \multicolumn{1}{r}{{32.4}} \\
\bottomrule
\end{tabular}
\begin{tablenotes}
      \item[a] $R_n$ is the mean absolute error (MAE) $R=\sqrt{\sum_i^N{\frac{(E^i-T^i)^2}{N}}}$ for each symmetric block (a symmetric block is a state whose state function differs only by the total angular momentum $J$).
\end{tablenotes}
\end{threeparttable}}
\end{table}

For the fitting of configuration $4f^{13}5d6s^2$ using different methods, we also analyzed the variation trend of the radial integral parameters of its Hamiltonian matrix elements (Table \ref{Tab2}). Different interactions between electrons are expressed for the six parameters included in its Hamiltonian. From the results given by the Cowan code in Fit I and II, we observed some unreasonable parameter values after fitting that are too large or equal to 0, which is the result of overfitting by the LSF. While the LSF modifies the \emph{ab initio} results to a certain extent, the physically meaningful Slater parameters are trimmed excessively. Compared with the radial integral parameters of the Hamiltonian matrix elements calculated by the Cowan code, the parameter results we give are more stable, avoiding the result that the integral term is equal to 0 or too large, and the deviation from the \emph{ab initio} results is not large. Parameter values are more physically meaningful.

\begin{table}[H]
\centering
    \caption{{Yb I} $4f^{13}6s^{2}6p$ fitting calculation parameter values}\label{Tab2}
\resizebox{\linewidth}{!}{
\begin{tabular}{lS[table-format=5.1]S[table-format=5.1]S[table-format=5.1]S[table-format=5.1]S[table-format=5.1]S[table-format=5.1]S[table-format=5.1]}
\toprule

    \multicolumn{1}{c}{} & \multicolumn{1}{c}{} & \multicolumn{2}{c}{Fit I}                                  & \multicolumn{2}{c}{Fit II}                                 & \multicolumn{2}{c}{Fit III}                                 \\
\cline{3-8}
\multicolumn{1}{c}{{Parameters}}                            & \multicolumn{1}{c}{\emph{Ab initio}}                           & \multicolumn{1}{c}{Cowan code} & \multicolumn{1}{c}{This work} & \multicolumn{1}{c}{Cowan code} & \multicolumn{1}{c}{This work} & \multicolumn{1}{c}{Cowan code} & \multicolumn{1}{c}{This work} \\
\midrule
\({E}_{{\text{av}}}\)                                              & 38565.6                                        & 39467.2                     & 38565.6                      & 39659.1                     & 38755.0                      & 38796.9                     & 38796.9                      \\
\({\zeta}_{{f}}\)                                              & 3111.8                                         & 4243.5                      & 3190.1                       & 2320.3                      & 2947.7                       & 2903.0                      & 2903.0                       \\
\({\zeta}_{{p}}\)                                              & 1514.7                                         & 625.1                       & 1552.8                       & 1170.4                      & 2141.1                       & 2136.2                      & 2136.2                       \\
\({F}^{{2}}\left( {\text{fp}} \right)\)                                              & 2396.6                                         & 2102.2                      & 2396.6                       & 24744.7                     & 2396.6                       & 4474.6                      & 4474.6                       \\
\({G}^{{2}}\left( {\text{fp}} \right)\)                                              & 566.1                                          & 6214.3                      & 580.3                        & 6412.3                      & 566.1                        & 1046.8                      & 1046.8                       \\
\({G}^{{4}}\left( {\text{fp}} \right)\)                                              & 500.8                                          & 1211.0                      & 513.4                        & 0.0                         & 1715.6                       & 1027.2                      & 1027.2\\
\bottomrule
\end{tabular}}
\end{table}

In addition, we also compared some configurations of single-electron excitations in the $6s$ shell under different methods. From Fit I, the results obtained by the ridge regression model and the least squares method used by Cowan can fully demonstrate the advantages of the ridge regression model in the case of fewer data as well. Adding the same amount of experimental data, the MAE of our improved results and NIST experimental values are an order of magnitude higher than Cowan code's accuracy. Importantly, more physically meaningful results have been obtained. For configuration $4f^{14}6s6d$, in Fit I, we add two experimental energy levels to fit the calculation results. According to the results calculated by the Cowan code, the two spectral terms $^1\!D_2$ and $^3\!D_3$ without experimental values are 61154.1 and 61110.0 \si{cm^{-1}}, respectively, which deviates from the experimental and \emph{ab initio} results by 2000 \si{cm^{-1}}. The reason is the over-fitting phenomenon caused by the least square method. In order to keep the fitted function more consistent with the results of the added experimental data, there is a large error in the extension. However, when the same data is added to the ridge regression model, the above situation does not occur, and the extrapolated results are more realistic, with a difference of less than 300 \si{cm^{-1}} from the experimental energy level. There is no doubt that the accuracy can also be greatly improved if more experimental data is included. It is only 0.9 \si{cm^{-1}} different from the experimental data for the same symmetric block.

\begin{table}[H]
\centering
    \caption{{Yb I} $4f^{14}6snl$ fitting calculated energy level values (in \si{\centi\metre^{-1}}). The data with asterisks (*) are the data fitted with the experimental values.}\label{Tab3}
\resizebox{\linewidth}{!}{
\begin{threeparttable}
\begin{tabular}{cccS[table-format=5.1]S[table-format=5.1]S[table-format=5.1]S[table-format=5.1]S[table-format=5.1]S[table-format=5.1]}
\toprule
\multicolumn{1}{c}{} &
  &
  &
  \multicolumn{1}{c}{} &
  \multicolumn{1}{c}{} &
  \multicolumn{2}{c}{Fit I} &
  \multicolumn{2}{c}{Fit II} \\ \cline{6-9} 
\multicolumn{1}{c}{{Configuration}} &
   {Term} &
   {$J$} &
  \multicolumn{1}{c}{{Expt.}} &
  \multicolumn{1}{c}{\emph{Ab initio}} &
  \multicolumn{1}{c}{Cowan code} &
  \multicolumn{1}{c}{This work} &
  \multicolumn{1}{c}{Cowan code} &
  \multicolumn{1}{c}{This work} \\ 
\midrule
    $4f^{14}6s^2$              & $^1\!S$ & 0 & 0.0     & 0.0     & 0.0* & 0.0* & 0.0       & 0.0       \\
    $4f^{14}6s6p$             & $^3\!P$ & 0 & 17288.4 & 11235.3 & 17269.9* & 17288.4* & 17288.4* & 17288.4* \\
                          &       & 1 & 17992.0 & 11668.9 & 17930.3  & 18009.0  & 17908.8   & 17908.8   \\
                          &       & 2 & 19710.4 & 12626.1 & 19924.4  & 20421.9  & 19710.4* & 19710.4* \\
                          & $^1\!P$& 1 & 25068.2 & 26035.3 & 25132.0* & 25068.5* & 25068.2* & 25068.2* \\
    \multicolumn{1}{c}{$R_1$\tnote{a}} &       &   &         & \multicolumn{1}{r}{5651.2}  & \multicolumn{1}{r}{116.2}  & \multicolumn{1}{r}{355.8}  & \multicolumn{1}{r}{32.2}      & \multicolumn{1}{r}{32.2}      \\
    $4f^{14}5d6s$             & $^3\!D$ & 1 & 24489.1 & 20789.3 & 24489.1* & 24489.1* & 24489.1*  & 24489.1*  \\
                          &       & 2 & 24751.9 & 21257.1 & 24780.3  & 24834.2  & 24751.9*  & 24751.9*  \\
                          &       & 3 & 25270.9 & 22128.8 & 27069.3  & 26865.7  & 25275.2   & 25275.2   \\
                          & $^1\!D$ & 2 & 27677.6 & 27658.7 & 27677.7* & 27678.5* & 27677.7*  & 27677.7*  \\
    \multicolumn{1}{c}{$R_2$} &       &   &         & \multicolumn{1}{r}{2990.7}  & \multicolumn{1}{r}{899.3}  & \multicolumn{1}{r}{798.4}  & \multicolumn{1}{r}{2.2}       & \multicolumn{1}{r}{2.2}       \\
    $4f^{14}6s7p$             & $^3\!P$ & 0 & 38090.7 & 33330.9 & 38090.7* & 38090.7* & 38067.6*  & 38067.6*  \\
                          &       & 1 & 38174.2 & 33421.3 & 38174.2* & 38174.2* & 38204.9*  & 38204.9*  \\
                          &       & 2 & 38552.0 & 33639.9 & 38350.2  & 38373.0  & 38544.6*  & 38544.6*  \\
                          & $^1\!P$& 1 & 40564.0 & 35109.1 & 43119.1  & 39849.0  & 40563.7*  & 40563.7*  \\
    \multicolumn{1}{c}{$R_3$} &       &   &         & \multicolumn{1}{r}{4978.2}  & \multicolumn{1}{r}{1281.5}  & \multicolumn{1}{r}{368.5}  & \multicolumn{1}{r}{19.6}      & \multicolumn{1}{r}{19.6}      \\
    $4f^{14}6s6d$             & $^3\!D$ & 1 & 39808.7 & 35330.3 & 39808.7* & 39808.7* & 39807.6*  & 39807.6*  \\
                          &       & 2 & 39838.0 & 35389.7 & 39838.0* & 39838.0* & 39839.3*  & 39839.3*  \\
                          &       & 3 & 39966.1 & 35526.8 & 61110.0  & 39887.1  & 39966.2*  & 39966.2*  \\
                          & $^1\!D$ & 2 & 40061.5 & 35893.0 & 61154.1  & 40571.3  & 40061.3*  & 40061.3*  \\
    \multicolumn{1}{c}{$R_4$} &       &   &         & \multicolumn{1}{r}{4385.4}  & \multicolumn{1}{r}{14932.9}  & \multicolumn{1}{r}{257.9}  & \multicolumn{1}{r}{0.9}       & \multicolumn{1}{r}{0.9}       \\
\bottomrule   
\end{tabular}
\begin{tablenotes}
      \item[a] $R_n$ is the mean absolute error (MAE) $R=\sqrt{\sum_i^N{\frac{(E^i-T^i)^2}{N}}}$ for each symmetric block (a symmetric block is a state whose state function differs only by the total angular momentum $J$).
\end{tablenotes}
\end{threeparttable}}
\end{table}

Based on the ridge regression model, in the case of few data, the \emph{ab initio} results can be modified to a certain extent for different excited states to optimize the Hamiltonian. We use this method to perform fitting calculations for other excited states, which are summarized in Table \ref{Tab4} and Table \ref{Tab5} (odd and even parity, respectively).

\begin{table}[H]
\centering
\caption{{Yb I} odd-parity excited state fitting calculated energy levels (in \si{\centi\metre^{-1}}), Landé $g$-factor values.}\label{Tab4}
\renewcommand{\arraystretch}{0.7}
\resizebox{\linewidth}{!}{
\begin{tabular}{cccS[table-format=5.1]S[table-format=5.1]S[table-format=3.1]S[table-format=1.5]S[table-format=1.3]S[table-format=1.5]}
\toprule
 &  &  & \multicolumn{2}{c}{Level}                             & \multicolumn{1}{c}{} & \multicolumn{2}{c}{Landé $g$}                           & \multicolumn{1}{c}{} \\ \cline{4-5} \cline{7-8}
 {Configuration}      &      {Term}    &        {$J$}            & \multicolumn{1}{c}{Expt.} & \multicolumn{1}{c}{This work} & \multicolumn{1}{c}{$\Delta E$}   & \multicolumn{1}{c}{Expt.} & \multicolumn{1}{c}{This work} & \multicolumn{1}{c}{{$\Delta g$}}                    \\ 
\midrule
    $4f^{14}6s6p$                       & $^3\!P$                    & 0                  & 17288.4                 & 17288.4                     & 0.0                    &                         & 0.000                       &                                         \\
                               &                       & 1                  & 17992.0                 & 17908.8                     & 83.2                   & 1.49282                 & 1.488                       & 0.00482                                 \\
                               &                       & 2                  & 19710.4                 & 19710.4                     & 0.0                    & 1.5                     & 1.501                       & 0.001                                   \\
                               & $^1\!P$                   & 1                  & 25068.2                 & 25068.2                     & 0.0                    & 1.035                   & 1.013                       & 0.022                                   \\
    $4f^{13}5d6s^2$                      & (7/2, 3/2)             & 2                  & 23188.5                 & 23339.9                     & 151.4                  & 1.45                    & 1.461                       & 0.011                                   \\
                               &                       & 5                  & 25859.6                 & 25852.5                     & 7.1                    & 1.04                    & 1.023                       & 0.017                                   \\
                               &                       & 3                  & 27445.6                 & 27309.2                     & 136.4                  & 1.22                    & 1.223                       & 0.003                                   \\
                               &                       & 4                  & 28184.5                 & 28150.4                     & 34.1                   & 1.14                    & 1.137                       & 0.003                                   \\
                               & (7/2, 5/2)             & 6                  & 27314.9                 & 27035.7                     & 279.2                  & 1.16                    & 1.167                       & 0.007                                   \\
                               &                       & 2                  & 28196.0                 & 28358.0                     & 162.0                  & 1.02                    & 1.023                       & 0.003                                   \\
                               &                       & 1                  & 28857.0                 & 28857.8                     & 0.8                    & 1.2635                  & 1.272                       & 0.0085                                  \\
                               &                       & 4                  & 29775.0                 & 29885.2                     & 110.2                  & 1.09                    & 1.091                       & 0.001                                   \\
                               &                       & 3                  & 30207.4                 & 30215.0                     & 7.6                    & 1.08                    & 1.091                       & 0.011                                   \\
                               &                       & 5                  & 30524.7                 & 30556.1                     & 31.4                   & 1.18                    & 1.176                       & 0.004                                   \\
                               & (5/2, 5/2)             & 0                  &                         & 36047.2                     &                        &                         & 0.000                       &                                         \\
                               &                       & 1                  &                         & 39156.9                     &                        &                         & 0.768                       &                                         \\
                               &                       & 5                  &                         & 39660.1                     &                        &                         & 1.035                       &                                         \\
                               &                       & 2                  &                         & 40499.2                     &                        &                         & 0.918                       &                                         \\
                               &                       & 3                  &                         & 41784.5                     &                        &                         & 0.979                       &                                         \\
                               &                       & 4                  &                         & 42349.4                     &                        &                         & 1.045                       &                                         \\
                               & (5/2, 3/2)             & 4                  &                         & 37291.3                     &                        &                         & 0.827                       &                                         \\
                               &                       & 2                  &                         & 38593.0                     &                        &                         & 0.933                       &                                         \\
                               &                       & 1                  &                         & 41961.5                     &                        &                         & 0.96                        &                                         \\
                               &                       & 3                  &                         & 40740.8                     &                        &                         & 0.874                       &                                         \\
    $4f^{14}6s7p$                       & $^3\!P$                    & 0                  & 38090.7                 & 38067.6                     & 23.1                   &                         & 0.000                       &                                         \\
                               &                       & 1                  & 38174.2                 & 38204.9                     & 30.7                   & 1.14                    & 1.497                       & 0.357                                   \\
                               &                       & 2                  & 38552.0                 & 38544.6                     & 7.4                    & 1.5                     & 1.501                       & 0.001                                   \\
                               & $^1\!P$                   & 1                  & 40564.0                 & 40563.7                     & 0.3                    & 1.01                    & 1.005                       & 0.005                                   \\
    $4f^{14}6s5f$                       & $^1\!F$                    & 3                  &                         & 42680.6                     &                        &                         & 1.045                       &                                         \\
                               & $^3\!F$                    & 4                  &                         & 42652.3                     &                        &                         & 1.251                       &                                         \\
                               &                       & 3                  &                         & 43456.6                     &                        &                         & 1.039                       &                                         \\
                               &                       & 2                  & 43433.9                 & 43433.8                     & 0.0                    & 0.68                    & 0.666                       & 0.014                                   \\
    $4f^{14}6s8p$                       & $^3\!P$                    & 0                  & 43614.3                 & 43616.9                     & 2.6                    &                         & 0.000                       &                                         \\
                               &                       & 1                  & 43659.4                 & 43656.4                     & 3.0                    & 1.48                    & 1.469                       & 0.011                                   \\
                               &                       & 2                  & 43805.7                 & 43805.9                     & 0.2                    & 1.49                    & 1.501                       & 0.011                                   \\
                               & $^1\!P$                   & 1                  & 44017.6                 & 44017.8                     & 0.2                    & 1                       & 1.033                       & 0.033                                   \\
    $4f^{14}6s11p$                      & (1/2, 1/2)             & 0                  &                         & 48192.7                     &                        &                         & 0.000                       &                                         \\
                               &                       & 1                  & 48212.1                 & 48198.6                     & 13.5                   &                         & 1.49                        &                                         \\
                               & (1/2, 3/2)             & 2                  &                         & 48215.5                     &                        &                         & 1.501                       &                                         \\
                               &                       & 1                  & 48258.5                 & 48271.9                     & 13.4                   &                         & 1.011                       &                                         \\
    $4f^{14}6s6f$                       & $^1\!F$                    & 3                  &                         & 45143.7                     &                        &                         & 1.000                       &                                         \\
                               & $^3\!F$                    & 2                  & 45956.3                 & 45956.3                     & 0.0                    & 0.72                    & 0.666                       & 0.054                                   \\
                               &                       & 3                  &                         & 46169.9                     &                        &                         & 1.084                       &                                         \\
                               &                       & 4                  &                         & 46360.1                     &                        &                         & 1.251                       &                                         \\
    $4f^{14}6s9p$                       & $^3\!P$                    & 0                  & 46082.2                 & 46068.5                     & 13.7                   &                         & 0.000                       &                                         \\
                               &                       & 1                  & 46078.9                 & 46095.4                     & 16.5                   & 1.34                    & 1.481                       & 0.141                                   \\
                               &                       & 2                  & 46184.2                 & 46182.0                     & 2.2                    & 1.5                     & 1.501                       & 0.001                                   \\
                               & $^1\!P$                    & 1                  & 46370.3                 & 46369.6                     & 0.7                    & 1.07                    & 1.02                        & 0.05                                    \\ 
\bottomrule 
\end{tabular}}
\end{table}

\begin{table}[H]
\centering
\caption{{Yb I} even-parity excited state fitting calculated energy levels (in \si{\centi\metre^{-1}}), Landé $g$-factor values}\label{Tab5}
\renewcommand{\arraystretch}{0.7}
\resizebox{\linewidth}{!}{
\begin{tabular}{cccS[table-format=5.1]S[table-format=5.1]S[table-format=2.1]S[table-format=1.2]S[table-format=1.3]S[table-format=1.3]}
\toprule
 &  &  & \multicolumn{2}{c}{Level}                             & \multicolumn{1}{c}{} & \multicolumn{2}{c}{Landé $g$}                           & \multicolumn{1}{c}{} \\ \cline{4-5} \cline{7-8}
 {Configuration}      &      {Term}    &        {$J$}            & \multicolumn{1}{c}{Expt.} & \multicolumn{1}{c}{This work} & \multicolumn{1}{c}{$\Delta E$}   & \multicolumn{1}{c}{Expt.} & \multicolumn{1}{c}{This work} & \multicolumn{1}{c}{{$\Delta g$}}                    \\ 
\midrule
$4f^{14}6s^2$              & $^1\!S$                   & {0}                    & 0.0            & 0.0               & 0.0                           &                 & 0.000              & \multicolumn{1}{l}{}          \\
$4f^{14}5d6s$             & $^3\!D$                   & {1}                    & 24489.1        & 24489.1           & 0.0                           & 0.5             & 0.499              & 0.001                         \\
                              & {}                     & {2}                    & 24751.9        & 24751.9           & 0.0                           & 1.16            & 1.164              & 0.004                         \\
                              & {}                     & {3}                    & 25270.9        & 25275.2           & 4.3                           & 1.34            & 1.334              & 0.006                         \\
                              & $^1\!D$                   & {2}                    & 27677.6        & 27677.7           & 0.1                           & 1.01            & 1.003              & 0.007                         \\
$4f^{14}6s7s$             & $^3\!S$                   & {1}                    & 32694.7        & 32694.7           & 0.0                           & 2.01            & 2.002              & 0.008                         \\
{}                     & $^1\!S$                   & {0}                    & 34350.6        & 34350.6           & 0.0                           &                 & 0.000              &                               \\
$4f^{14}6s6d$             & $^3\!D$                   & {1}                    & 39808.7        & 39807.6           & 1.1                           & 0.5             & 0.499              & 0.001                         \\
\multicolumn{1}{l}{}          & {}                     & {2}                    & 39838.0        & 39839.3           & 1.3                           & 1.16            & 1.143              & 0.017                         \\
\multicolumn{1}{l}{}          & {}                     & {3}                    & 39966.1        & 39966.2           & 0.1                           & 1.33            & 1.334              & 0.004                         \\
                              & $^1\!D$                   & {2}                    & 40061.5        & 40061.3           & 0.2                           & 1.03            & 1.024              & 0.006                         \\
$4f^{13}6s^26p$            & {(7/2, 1/2)}            & {3}                    & 32065.3        & 32069.0           & 3.7                           & 1.23            & 1.259              & 0.029                         \\
{}                     & {}                     & {4}                    & 32273.6        & 32280.3           & 6.7                           &                 & 1.063              &                               \\
{}                     & {(7/2, 3/2)}            & {5}                    & 35178.8        & 35134.8           & 44.0                          &                 & 1.200              &                               \\
{}                     & {}                     & {2}                    & 35197.0        & 35205.4          & 8.4                           & 1.05            & 1.057              & 0.007                         \\
{}                     & {}                     & {3}                    & 35807.5        & 35789.2           & 18.3                          & 1.08            & 1.088              & 0.008                         \\
{}                     & {}                     & {4}                    & 36061.0        & 36104.5           & 43.5                          &                 & 1.200              &                               \\
{}                     & {(5/2, 1/2)}            & {3}                    &                & 42343.4           &                               &                 & 0.799              &                               \\
{}                     & {}                     & {2}                    & 42531.9        & 42521.9           & 10.0                          &                 & 0.965              &                               \\
{}                     & {(5/2, 3/2)}            & {3}                    &                & 46218.5           &                               &                 & 1.021              &                               \\
{}                     & {}                     & {1}                    & 44834.6        & 44877.6           & 43.0                          & 0.66            & 0.499              & 0.161                         \\
{}                     & {}                     & {4}                    & 45497.6        & 45473.3           & 24.3                          &                 & 1.037              &                               \\
{}                     & {}                     & {2}                    & 45931.9        & 45905.1           & 26.8                          &                 & 0.811              &                               \\
$4f^{14}6s8s$             & $^3\!S$                   & {1}                    & 41615.0        & 41615.0           & 0.0                           & 2.02            & 2.002              & 0.018                         \\
\multicolumn{1}{l}{}          & $^1\!S$                   & {0}                    & 41939.9        & 41939.9           & 0.0                           &                 & 0.000              &                               \\
$4f^{14}6s7d$             & $^3\!D$                   & {1}                    & 44311.4        & 44302.3           & 9.1                           &                 & 0.499              &                               \\
\multicolumn{1}{l}{}          & {}                     & {2}                    & 44313.1        & 44314.1           & 1.0                           &                 & 1.157              &                               \\
\multicolumn{1}{l}{}          & {}                     & {3}                    & 44357.6        & 44345.3           & 12.3                          & 1.32            & 1.334              & 0.014                         \\
\multicolumn{1}{l}{}          & $^1\!D$                   & {2}                    & 44380.8        & 44401.1           & 20.3                          & 1.1             & 1.010              & 0.09                          \\
$4f^{14}6p^2$              & $^3\!P$                   & {0}                    & 42436.9        & 42447.5           & 10.6                          &                 & 0.000              &                               \\
\multicolumn{1}{l}{}          & \multicolumn{1}{l}{}          & {1}                    & 43805.4        & 43763.7           & 41.7                          & 1.47            & 1.501              & 0.031                         \\
\multicolumn{1}{l}{}          & \multicolumn{1}{l}{}          & {2}                    & 44760.4        & 44806.1           & 45.7                          & 1.34            & 1.436              & 0.096                         \\
\multicolumn{1}{l}{}          & $^1\!D$                   & {2}                    & 47821.8        & 47819.8           & 2.0                           & 1.04            & 1.065              & 0.025                         \\
\multicolumn{1}{l}{}          & $^1\!S$                   & {0}                    &                & 49935.2           &                               &                 & 0.000              &                               \\
$4f^{14}5d^2$              & $^3\!F$                   & {2}                    & 47634.4        & 47634.4           & 0.0                           &                 & 0.668              &                               \\
\multicolumn{1}{l}{}          & \multicolumn{1}{l}{}          & {3}                    & 47860.2        & 47860.3           & 0.1                           & 1.02            & 1.084              & 0.064                         \\
\multicolumn{1}{l}{}          & \multicolumn{1}{l}{}          & {4}                    &                & 48130.4           & \multicolumn{1}{l}{}          &                 & 1.25               &                               \\
\multicolumn{1}{l}{}          & $^1\!D$                   & {2}                    &                & 50036.9           & \multicolumn{1}{l}{}          &                 & 1.048              &                               \\
\multicolumn{1}{l}{}          & $^3\!P$                   & {0}                    &                & 50404.0           & \multicolumn{1}{l}{}          &                 & 0.000              &                               \\
\multicolumn{1}{l}{}          & \multicolumn{1}{l}{}          & {1}                    &                & 50492.5           & \multicolumn{1}{l}{}          &                 & 1.501              &                               \\
\multicolumn{1}{l}{}          & \multicolumn{1}{l}{}          & {2}                    &                & 50697.6           & \multicolumn{1}{l}{}          &                 & 1.45               &                               \\
\multicolumn{1}{l}{}          & $^1\!G$                   & {4}                    &                & 51319.5           & \multicolumn{1}{l}{}          &                 & 1.000              &                               \\
\multicolumn{1}{l}{}          & $^4\!S$                   & {0}                    &                & 56283.8           & \multicolumn{1}{l}{}          &                 & 0.000              &                               \\
$4f^{14}6s8d$             & $^3\!D$                   & {1}                    & 46445.0        & 46451.6           & 6.6                           & 0.49            & 0.499              & 0.009                         \\
\multicolumn{1}{l}{}          & {}                     & {2}                    & 46467.7        & 46459.7           & 8.0                           & 1.12            & 1.156              & 0.036                         \\
\multicolumn{1}{l}{}          & {}                     & {3}                    & 46480.7        & 46482.0           & 1.3                           & 1.35            & 1.334              & 0.016                         \\
\multicolumn{1}{l}{}          & $^1\!D$                   & {2}                    &                & 46519.5           &                               &                 & 1.011              &    \\
\bottomrule                          
\end{tabular}}
\end{table}

\section{Conclusion}
For the {Yb I} atom, we propose and use the data-driven method of ridge regression model fitting to calculate a total of 94 energy levels for the even-parity configuration $4f^{14}6s^{2}$ + $4f^{14}5d6s$ + $4f^{14}6s7s$ + $\ldots{}$ and the odd-parity group $4f^{14}6s6p$ + $4f^{13}5d6s^{2}$ + $4f^{14}6s7p$ + $\ldots{}$, where there are 66 experimental energy levels added, and these experimental energy levels fully consider the electron correlation. The $4f$ shell is full of electrons, and the absolute error between the calculated energy level and the NIST experimental value is less than 50 \si{cm^{-1}}, and the relative is less than 0.1\%. Among them, the $4f^{13}5d6s^{2}$ configuration has an average absolute error of 70 \si{cm^{-1}} and a relative error of less than 0.3\%. Moreover, reliable predicted data are made for energy levels that are experimentally unattainable. This work demonstrates that the ridge regression model has a stronger generalization ability when the amount of data is insufficient, and the predicted data is more reliable. It is expected that a more accurate machine learning model will be applied to the atomic structure calculation.


\addcontentsline{toc}{chapter}{References}
\footnotesize

\begin{thebibliography}{99}\footnotesize

\bibitem{ning_recent_2022}
Ning Y, Jin G~Q, Wang M~X, Gao S and Zhang J~L 2022 \emph{Curr. Opin. Chem.
  Biol.} {\bf 66} 102097

\bibitem{cisek_wide-field_2017}
Zhao H, Cisek R and Barzda V 2017 \emph{Photonics North (PN)} June  6--8, 2017, Ottawa, Canada, p. 1

\bibitem{ludlow_optical_2015}
Ludlow A~D, Boyd M~M, Ye J, Peik E and Schmidt P~O 2015 \emph{Rev. Mod. Phys.}
  {\bf 87} 637

\bibitem{eshkabilov_laser_2022}
Eshkabilov N~B, Kurbaniyazov A~S and Haidarov S~R 2022 \emph{Russ. Phys. J.} {\bf 64}
  1872

\bibitem{heugel_resonant_2016}
Heugel S, Fischer M, Elman V, Maiwald R, Sondermann M and Leuchs G 2016
  \emph{J. Phys. B: At. Mol. Opt.} {\bf 49} 015002

\bibitem{sahoo_investigations_2021}
Sahoo A~C, Mandal P~K, Mukherjee J, Dev V and Shah M~L 2021 \emph{J. Quant.
  Spectrosc. Radiat. Transf.} {\bf 276} 107944

\bibitem{xu_study_2018}
Xu Y~W, Shen L and Dai C~J 2018 \emph{Mod. Phys. Lett. B} {\bf 32} 1850190

\bibitem{galindo-uribarri_high_2021}
Galindo-Uribarri A, Liu Y, Romero~Romero E and Stracener D~W 2021 \emph{Sci.
  Rep.} {\bf 11} 23432

\bibitem{furmann_experimental_2014}
Furmann B and Stefanska D 2014 \emph{Phys. Scr.} {\bf 89} 095402

\bibitem{kneip_highly_2020}
Kneip N, Düllmann C~E, Gadelshin V, Heinke R, Mokry C, Raeder S, Runke J,
  Studer D, Trautmann N, Weber F and Wendt K 2020 \emph{Hyperfine Interact.}
  {\bf 241} 45

\bibitem{block_direct_2019}
Block M 2019 \emph{Radiochimica Acta} {\bf 107} 821

\bibitem{kramida_nist_2022}
Kramida A, Ralchenko Y, Reader J and NIST ASD Team 2021 \emph{NIST Atomic Spectra Database} (version 5.9), [Online]. Available: https://physics.nist.gov/asd [Sat Jul 09 2022]. National Institute of Standards and Technology, Gaithersburg, MD. DOI: https://doi.org/10.18434/T4W30F

\bibitem{cowan_theoretical_1968}
Cowan R~D 1968 \emph{J. Opt. Soc. Am.} {\bf 58} 808

\bibitem{froese_fischer_mchf_2007}
Froese~Fischer C, Tachiev G, Gaigalas G and Godefroid M~R 2007 \emph{Comput.
  Phys. Commun.} {\bf 176} 559
  
\bibitem{hibbert_civ3_1975}
Hibbert A 1975 \emph{Comput. Phys. Commun.} {\bf 9} 141

\bibitem{jonsson_grasp2k_2013}
Jönsson P, Gaigalas G, Bieroń J, Fischer C~F and Grant I 2013 \emph{Comput.
  Phys. Commun.} {\bf 184} 2197

\bibitem{gu_fac_2008}
Gu M~F 2008 \emph{Can. J. Phys.} {\bf 86} 675

\bibitem{MDFGME1}
Desclaux J P 1975 \emph{Comput. Phys. Commun.} {\bf 9} 31

\bibitem{MDFGME2}
Indelicato P, Gorceix O and Desclaux J P 1987  \emph{J. Phys. B: At. Mol. Opt.} {\bf 20} 651

\bibitem{MDFGME3}
Indelicato P and Desclaux J P 1990 \emph{Phys. Rev. A} {\bf 42} 5139

\bibitem{raassen_use_1996}
Raassen A~J~J and Uylings P~H~M 1996 \emph{Phys. Scr.} {\bf 1996} 84

\bibitem{hoyt_observation_2005}
Hoyt C~W, Barber Z~W, Oates C~W, Fortier T~M, Diddams S~A and Hollberg L 2005
  \emph{Phys. Rev. Lett.} {\bf 95} 083003

\bibitem{king_temperature_1931}
King A~S 1931 \emph{Astrophys. J.} {\bf 74} 328

\bibitem{meggers_iii_1942}
Meggers W~F 1942 \emph{Rev. Mod. Phys.} {\bf 14} 0097

\bibitem{niyaz_microwave_2019}
Niyaz F, Nunkaew J and Gallagher T~F 2019 \emph{Phys. Rev. A} {\bf 99}  042507

\bibitem{ternovsky_advanced_2018}
Ternovsky V~B, Kuznetsova A~A, Ternovsky E~V, Mironenko D~A and Smirnov A~V 2018 \emph{J. Phys.: Conf. Ser.} {\bf 1136} 012010

\bibitem{dzuba_fast_2019}
Dzuba V~A, Flambaum V~V and Kozlov M~G 2019 \emph{Phys. Rev. A} {\bf 99} 032501

\bibitem{hoerl_ridge_1970}
Hoerl A~E and Kennard R~W 1970 \emph{Technometrics} {\bf 12} 55

\bibitem{cowan_theory_1981}
Cowan R~D 1981 \emph{The Theory of Atomic Structure and Spectra} (Univ. of
  California Press)

\end{thebibliography}

\end{document}